\documentclass[twocolumn,prl,showpacs,preprintnumbers,amsmath,amssymb]{revtex4}
\usepackage{graphicx}
\begin{document}
\title{Electron-phonon interaction at the Be(0001) surface.}
\preprint{APS/123-QED}
\author{A. Eiguren$^{1}$, S. de Gironcoli$^{2}$, E. V. Chulkov$^{1,3}$, 
P. M. Echenique$^{1,3}$, and E. Tosatti$^{2,4}$ }
\address{1. Departmento de F\'{\i}sica de Materiales and Centro Mixto
  CSIC-UPV/EHU, Facultad de Ciencias Qu\'{\i}micas,
Universidad del Pais Vasco/Euskal Herriko Unibertsitatea, Adpo. 1072,
20018 San Sebasti\'{a}n/Donostia, Basque Country, Spain}
\address{2. International School for Advanced Studies (SISSA) 
and Istituto Nazionale di Fisica della Materia 
(INFM/DEMOCRITOS), via Beirut 2-4, I-34014, Trieste, Italy}
\address{3. Donostia International Physics Center (DIPC),~Paseo~de~Manuel~Lardizabal,~4,~20018~San~Sebasti\'{a}n/Donostia,~Spain}
\address{4. International Centre for Theoretical Physics (ICTP), 
P.O. Box 586, I-34014, Trieste, Italy}
%
%
\date{\today}
\begin{abstract}
We present a {\it first principle} 
study of the electron-phonon (e-p) interaction 
at the Be(0001) surface. 
The real and imaginary part of the e-p self 
energy ($\Sigma$) 
are calculated for the $\bar{\Gamma}$ surface
state in the binding energy range from the 
$\bar{\Gamma}$ point to the Fermi level. 
Our calculation shows an overall good agreement with 
several photoemission 
data measured
at high and low temperatures. Additionally, 
we show that the energy derivative of $Re\Sigma$
presents a strong temperature and energy
variation close to $E_{F}$, making
it difficult to measure its value just at $E_{F}$.
\end{abstract}
\pacs{71.10.-w, 68.35.Ja, 73.20.-r, 74.78.-w}
\maketitle

\begin{figure}
\includegraphics[width=0.4 \textwidth,height=0.35 \textwidth]{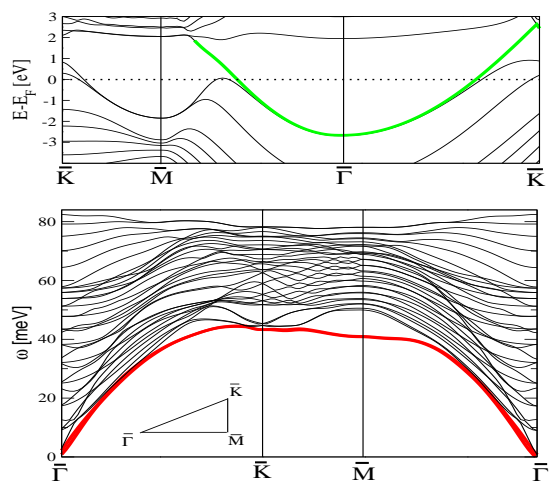}
\caption{\label{fig:bands} Top panel shows the electronic band structure
of a fully relaxed 12 layer Be(0001) slab, 
the $\bar{\Gamma}$ surface state is indicated by the green line.
Bottom panel shows the phonon dispersions where the red line
represents the Rayleigh mode.} 
\end{figure}
%
For many years now a considerable effort has been directed toward 
the study of 
electron-phonon (e-p) effects and quasiparticle dynamics in low dimensional systems.
Important examples include semiconductor heterostructures, 
superconducting materials with layered structures 
and electrons in two-dimensional surface states.
Experimentally, e-p effects are becoming increasingly accessible on surfaces,
through powerful techniques like high resolution angular-resolved 
photoemission spectroscopy (PES), 
scanning tunnelling spectroscopy (STS),
and time-resolved two-photon photoemission spectroscopy (TR-2PPE).

Be(0001) offers in this respect a textbook study case.
Here the surface states form a fully metallic, high density two-dimensional 
electron gas, rather well decoupled from the bulk Be substrate, 
itself only of semi-metal character \cite{Bart85,Chul87}. The surface states
increase the density of electron states (DOS) at the Fermi energy ($E_F$) by 
a factor of 4-5 in the surface layer \cite{Chul87}.
This radical change of character between 
surface and bulk goes along with some other very special 
properties of the Be(0001) surface \cite{general}. 
Following the DOS arguments, the surface $\lambda$ parameter, representative 
of the e-p interaction, is believed to be about a factor of 4-5 larger than 
in bulk Be \cite{Balasu1,Hengsberger-prl,Balasu2,Hengsberger-prb}. 
Values of $\lambda$ ranging from 0.65 to 1.2~ were obtained
from the high temperature surface state broadening \cite{Balasu1}, 
or by fitting the real part of the low temperatures surface
state self-energy to simple models \cite{Hengsberger-prl,Balasu2,Hengsberger-prb}.
\begin{figure}
\includegraphics[width=0.45 \textwidth,height=0.40\textwidth]{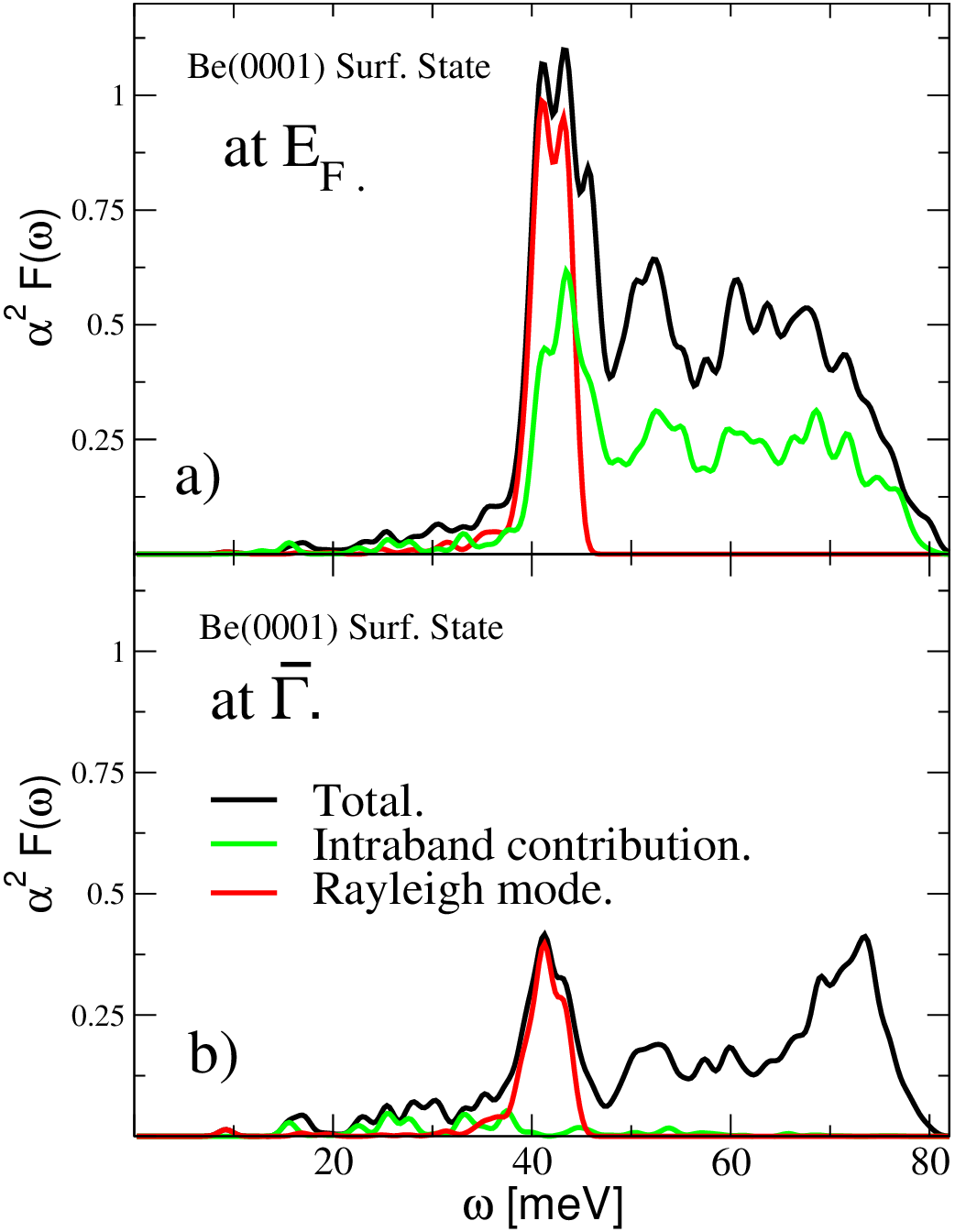}
\caption{\label{fig:Elias}  Different contributions to the Eliashberg
function for the hole state a) at $E_{F}$, b).
at $\bar{\Gamma}$ point.}
\end{figure}
The scatter between these values is disturbing. Also,
a value as large as 1.2 may raise questions concerning the stability
of the normal metallic state of this surface. 
On the other hand, it is well known that simple model calculations, 
such as Debye models, are surely too crude,
neglecting band structure effects and 
the important contribution of
the surface phonon modes \cite{Eiguren}. 
Electron energy-loss spectroscopy (EELS) measurements \cite{EELS2}, 
showed that the surface phonon dispersions 
at Be(0001) do qualitatively differ from those obtained by using truncated 
bulk models \cite{EELS2}.

In this letter we present the first  parameter free {\it ab-initio} 
calculation of the e-p interaction at a real metal surface, Be(0001),
that contains all the ingredients entering the e-p 
interaction, namely, surface electron and phonon states,
as well as electronic screening.
Our detailed analysis results in excellent agreement with 
all experimental data. We also motivate more future experimental work.
\begin{figure}
\includegraphics[width=0.45\textwidth,height=0.40 \textwidth]{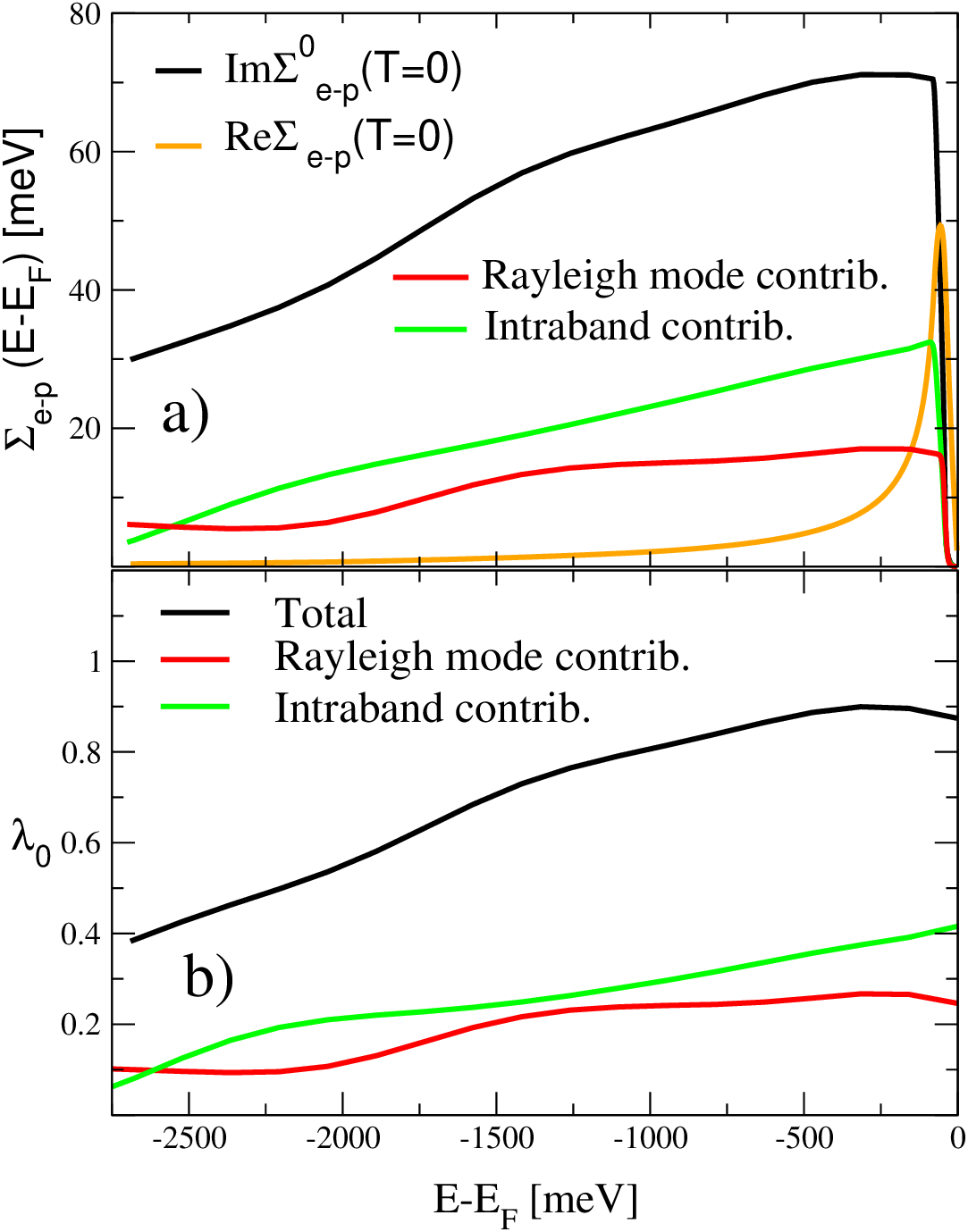}
\caption{\label{fig:ReS-ImS} {\bf a)} 
The e-p contribution to $Re\Sigma$ (orange)
and $Im\Sigma^0$ (black) at $T=0$ as function
of binding energy.
$Im\Sigma^0$ is 
broken up into contributions from the intraband 
scattering (green) and contribution from the Rayleigh mode (red).
{\bf b)} Black line represents the total value of 
$\lambda_{0}$ (Eq. \ref{landa}) and green and red lines the 
contributions from intraband scattering and Rayleigh mode, respectively.} 
\end{figure}
%

A fully relaxed 12 layer slab has been considered. 
We have used the density-functional theory in the local 
density approximation (LDA).
The e-p matrix elements and  
phonon frequencies were calculated using 
density functional perturbation theory \cite{Baroni}, through 
the {\tt PWSCF} program \cite{pwscf}.
A norm conserving pseudopotential was used with
a cutoff of $22 Ry$ for the expansion of the 
electronic wave function in plane waves.
For the Brillouin zone electronic structure  
integrations an order one Hermite-Gauss smearing technique 
was used, with smearing parameter $\sigma$ = $0.05$ Ry and as many as 
102 special {\bf k} points in the irreducible wedge of the $2D$ Brillouin zone (IWBZ).
The linear response calculation
for phonon modes  \cite{Baroni} have been performed with
30 special q points in the IWBZ, yielding surface phonons in 
excellent agreement with experiments \cite{EELS2} and previous calculations 
\cite{Gironcoli2}.

The basic quantity in the study of the e-p interaction 
is the so called Eliashberg function (ELF)~\cite{Grimvall81}. This function measures 
the phonon density of states weighted by the e-p interaction.
It represents the probability of emitting a phonon
of energy $\omega$ at $T=0$
\begin{eqnarray}
\alpha^2 F_{i}(\omega,E)=
\int \frac{d^2\vec{k}}{{\Omega_{BZ}}}\frac{\delta(E-E_{i,\vec{k}})}{N_{i}(E)} 
\times
\ \ \ \ \ \ \ \ \
\nonumber \\
\times
\int \frac{d^2\vec{q}}{\Omega_{BZ}} 
\sum^{f}_{\nu} \
|g_{\vec{k},\vec{k}+\vec{q}}^{i,f,\nu}|^2 \ 
\delta(E-E_{f,\vec{k}+\vec{q}})
\delta(\omega-\omega^{\nu}_{\vec{q}}).
\ \ \ 
\hfill\label{eq:Eli}
\end{eqnarray}
The ELF contains 
all the information needed for the calculation of the 
real and imaginary part of the self-energy. 
In Eq. \ref{eq:Eli},~$g_{\vec{k},\vec{k}+\vec{q}}^{i,f,\nu}\!\!=\!\!\langle \Psi_{i,\vec{k}}| 
\Delta_{SCF}V^{\nu}_{\vec{q}} |\Psi_{f,\vec{k}+\vec{q}} \rangle$ is 
the e-p matrix element
where $\Psi$ are the electronic state wave functions and 
$\Delta_{SCF} V^{\nu}_{\vec{q}}$ denotes the change of the 
self-consistent electron potential induced by a phonon mode with 
momentum $\vec{q}$ and mode index $\nu$;
$i$ refers to the electronic state (surface state) index, $f$ to the final state, 
$E$ denotes the binding energy, $N_{i}(E)$ the $i$ state partial density 
of states at energy $E$ and $\Omega_{BZ}$ is the area of the 2D Brillouin zone.

The imaginary part of the self-energy related to the 
lifetime of the excited electronic state through 
$\tau^{-1}=2 \ Im\Sigma$
can be written in terms of the ELF, 
\begin{eqnarray}\label{ImS}
Im\Sigma^0_{i}(E;T)= \pi
\int_{0}^{\infty}\!\!\!d\omega \
\alpha^2 F_{i}(\omega,E)
\nonumber \\
(1-f(E-\omega)+f(E+\omega)+2n(\omega))
,\end{eqnarray}
where $f$ and $n$ are the Fermi and Bose distribution functions, respectively.
The real part reflects renormalization of the electronic
band due to the interaction with phonons and can also be calculated from 
the ELF \cite{Grimvall81},  
\begin{eqnarray}
Re\Sigma_{i}(E;T) = 
\int^{\infty}_{-\infty}\!\!\!d\nu
\int_{0}^{\infty}\!\!\!d\omega
\ \alpha^2 F_{i}(\omega,E)
\nonumber \\
\frac{2\omega}{\nu^2-\omega^2} f(\nu+E)
.\end{eqnarray}
For the calculation of the ELF,
the {\bf k} integration was carried out with 102 special points
in the IWBZ and with 30 points for the {\bf q} integration. 
The two Dirac deltas containing the electron band energies 
were again replaced by
order one Hermite-Gauss smearing functions.
In order to gain confidence in the convergence of the calculation,
we performed a set of calculations with different
smearing parameters in the range $0.0025-0.1 Ry$, which gives an estimated
error of $\sim 5 \ \%$ for the surface state broadening at all energies  
except inside a window of $0.4 \ eV$ around $E_{F}$ where we 
estimate an error of $\sim 15 \ \%$.
\begin{figure}[h!!]
\includegraphics[width=0.39\textwidth,height=0.40\textwidth]{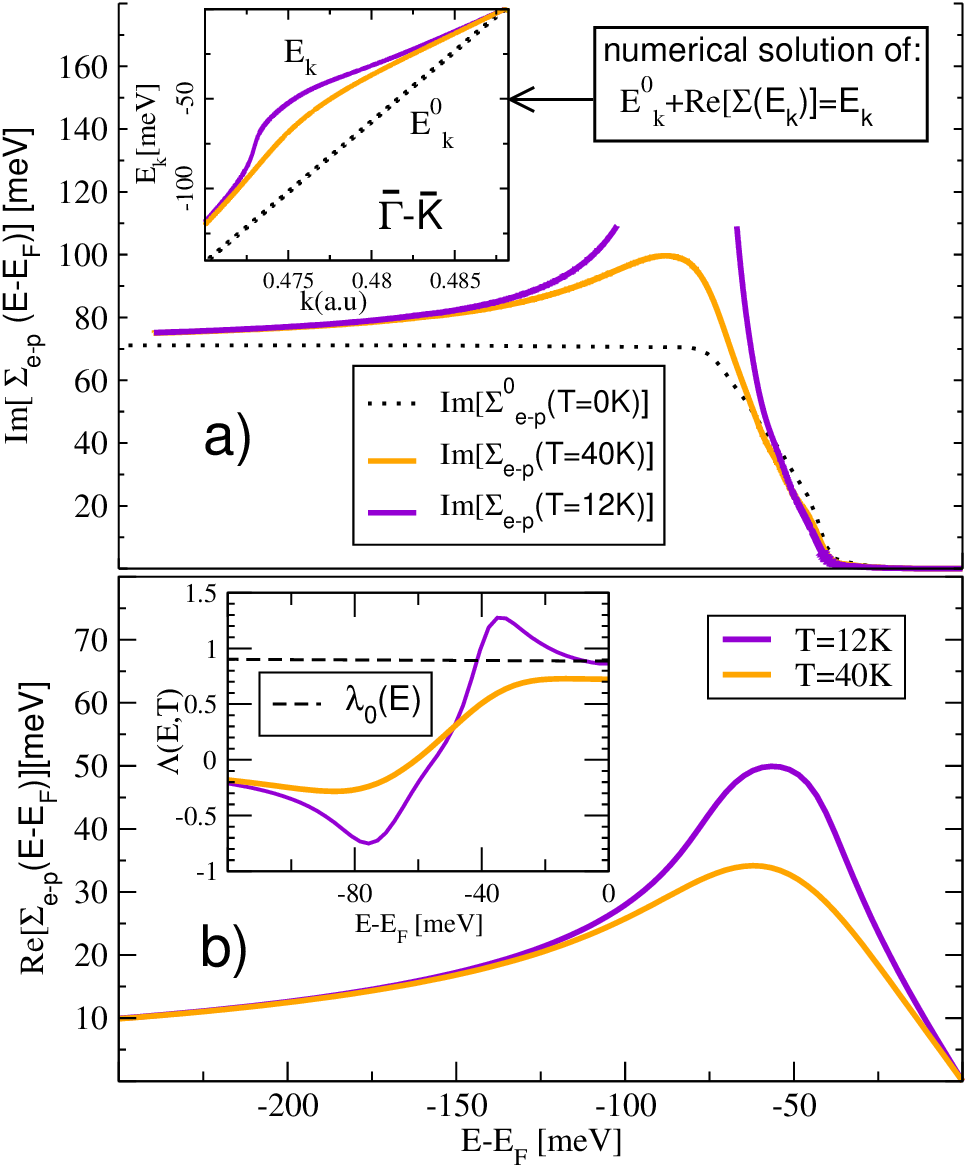}
\caption{\label{fig:Renor_ImS} 
{\bf a)} Surface state imaginary part of 
the self-energy near $E_{F}$. $Im\Sigma^0$ corresponds
to the unrenormalised result (Fig. \ref{fig:ReS-ImS}),
the renormalised $Im\Sigma$ is represented for temperatures
$12K$ (violet) and $40K$ (orange). The inset shows modified dispersions 
at $12K$ and $40K$. {\bf b)} $Re\Sigma$ near $E_{F}$ at
$12K$ and $40K$ (orange). The inset shows the quantity 
$\Lambda(E,T) \equiv -\frac{\partial \Sigma(E,T)}{\partial E}$.     
Note that at low temperature (12K) $\Lambda(E_{F},T) \simeq 0.9$ is
consistent with  $\lambda_{0}(E_{F})\simeq 0.9$ from Fig. \ref{fig:ReS-ImS}.}
\end{figure}

Fig. \ref{fig:bands} shows the electron and phonon band structures.
In the upper panel of Fig. \ref{fig:bands} the $\bar{\Gamma}$ 
point surface state band is sticked out (green line), near the 
$\bar{M}$ point another surface electronic state also appears.
In the lower panel, the phonon dispersions
are shown. The Rayleigh wave (red line) is a surface localised
phonon mode polarised mainly along the surface normal.
This and other surface phonon modes are visible 
within the surface band gaps.

Fig. \ref{fig:Elias} shows the calculated ELF
for a hole at $E_{F}$ (upper panel) 
and at the $\bar{\Gamma}$ point (lower panel).
The ELF is broken up into contributions
corresponding to the intraband
scattering (green line) and the Rayleigh mode scattering contribution (red line).  
From the upper panel of Fig. \ref{fig:Elias}, the decay of the hole 
at $E_F$ is dominated by the intraband scattering 
contribution (green line). The sum of the contributions from the decay into $\bar{M}$
point surface state (not represented) and the intraband contribution (green line)  
gives $\sim 90 \%$ of the total (solid line) at $E_F$. 
Thus, the decay process
of the hole at $E_{F}$ is sharply confined in space to
almost exclusively the surface state region.
By contrast, the character of the interaction changes radically 
for the hole at the $\bar{\Gamma}$ point (lower panel),
for which the decay into bulk states dominates. 

Photoemission experiments enable the measurement of the high 
temperature behaviour of the broadening, 
$\Gamma(E)= 2 \pi k_{B} T \lambda_{0}(E)$, for a state
with arbitrary binding energy $E$ \cite{Balasu1,Eiguren}. 
Theoretically $\lambda_{0}(E)$ is determined by the first 
reciprocal moment of the ELF~\cite{Grimvall81}. 
\begin{eqnarray}\label{landa}
\lambda_{0}(E) \equiv 2 \int_{0}^{\infty}\!\!\!
d \omega \ \frac{\alpha^2 F(\omega,E)}{\omega}
\end{eqnarray}
Eq. \ref{landa} indicates 
an enhanced role 
of the ELF at low energies.
Fig. \ref{fig:Elias} shows that the weight of the surface phonon
modes is concentrated around $\omega_{D}/2$ ($\omega_{D} \simeq 80 \ meV$),
thus a fitting procedure of $\lambda_{0}$ parameter through a 
Debye model ($\alpha^2F=\lambda_{0} (\omega/\omega_{D})^2$) will 
give rise in this case to an 
excessive low temperature broadening compared to a 
more realistic calculation using the same 
Debye energy $\omega_{D}$ and $\lambda_{0}$. 
The top panel of Fig. \ref{fig:ReS-ImS} shows the real (orange) and
imaginary part (black) of the self-energy calculated for T=0 and
as function of the binding energy. 
The imaginary part is resolved 
into contributions coming from intraband scattering (green) and
the contribution of the Rayleigh mode (red).
The imaginary part varies by a factor of two going from 
$E_{F}$ to the $\bar{\Gamma}$ point.
The bottom panel of Fig. \ref{fig:ReS-ImS} shows the variation of the 
$\lambda_{0}$ parameter defined through Eq. \ref{landa} as function 
of the binding energy. From this figure we deduce that 
the temperature derivative of the high-T surface
state broadening is twice for the hole 
at the $E_{F}$ compared to that at the $\bar{\Gamma}$ point.

Recently, the identity between the partial derivative of $Re\Sigma$ 
at Fermi level and $\lambda_{0}(E_{F})$, that is 
$\lambda \equiv \lambda_{0}(E_{F})=\Lambda(E_{F},T=0)$,
\begin{eqnarray}
\Lambda(E,T) \equiv - \frac{\partial Re\Sigma(E,T)}{\partial E}
\ \ \
\end{eqnarray}
has been used to determine the e-p coupling parameter $\lambda$ \cite{Hengsberger-prl,Balasu2,Hengsberger-prb}.
However, it is of crucial importance here 
to distinguish $\Lambda$ and $\lambda_{0}$.
The results of Fig. \ref{fig:ReS-ImS} do not
take into account the renormalization of electron bands
due to the real part of the self-energy. 
We denote by $E^{0}_{k}$ the pure LDA unrenormalised 
bands and by $E_{k}$ the renormalised ones including $Re\Sigma$, 
obtained by solving numerically the equation 
$E_{k}=E^{0}_{k}+Re\Sigma(E_{k})$~\cite{Grimvall81} (see Fig. \ref{fig:Renor_ImS}).
It is well known that close to $E_F$ 
the electron velocity, e-p coupling
form factor and the lifetime itself are renormalised \cite{Grimvall81}. 

Provided that the system is isotropic and $|\Lambda(E,T)|<<1$,
an estimation of the 
renormalised imaginary part \cite{Grimvall81} 
can be obtained from 
\begin{eqnarray}\label{renorm}
Im\Sigma(E_{k}) 
\simeq \frac{Im\Sigma^{0}(E_{k})}{1+\Lambda(E_{k})} 
,\end{eqnarray}
where $Im\Sigma$ denotes the renormalised quantity.
While the breakdown of the approximation expressed by Eq. \ref{renorm} is clear
for values of $\Lambda(E,T)$ close to -1, this form generally
helps us to understand how the deformation of the electronic
band caused by the real self-energy affects
itself the lifetime of the surface state. 

In the top panel of Fig. \ref{fig:Renor_ImS}, we show the 
unrenormalised imaginary part of the self-energy 
($Im\Sigma^0$, dotted line) calculated from  
Eq. \ref{ImS} and the renormalised ones at $T=40$ (orange) and 
at $T=12K$ (violet). 
The result for $12K$ is displayed only for reasonable 
values of $1/(1+\Lambda)$.  
These two temperatures, $12K$ and $40K$, are
the same as those considered experimentally, Refs.
\cite{Hengsberger-prl} and \cite{Balasu2}, respectively. 
In the bottom panel, the real part of the 
self-energy at $T=12K$ (violet) and at $T=40K$ (orange) are depicted.
The inset of the bottom panel shows the derivatives of the real part of the
self-energy at these two temperatures.

In Ref. \cite{Balasu2} a detailed measurement of 
$Im\Sigma$ and $Re\Sigma$
is reported for $T=40K$. An 'unusual' peak of $Im\Sigma$
for the hole at binding energies around
the maximum phonon energy ($\sim \omega_{D}$) 
was observed. This behaviour is very well
described by our calculation at the same temperature (orange).
The calculated real part of the self-energy also shows excellent
agreement with the measurement, and our 
$\Lambda(E_{F},T=40K) \simeq 0.7$, is in excellent agreement
with their measured $0.7 \pm 0.1$ 
(Inset of the bottom panel of Fig. \ref{fig:Renor_ImS}). 
In Ref. \cite{Hengsberger-prl,Hengsberger-prb} the derivative of
the real self-energy part 
was presented for $T=12 \ K$ suggesting a value of 
$\Lambda(E_{F},T=12K) \simeq 1.15$.
As explained  in Ref. \cite{Balasu2}, this value was slightly overestimated 
due to the fixed angle photoemission 
experiment, and a corrected value of $0.87$ was suggested instead.
Our calculated  $\Lambda(E_{F},T=12K) \simeq 0.9$ is in excellent agreement
with this corrected value
(inset of the bottom panel of Fig. \ref{fig:Renor_ImS}).
\begin{figure}[hb]
\includegraphics[width=0.35 \textwidth,height=0.2075\textwidth]{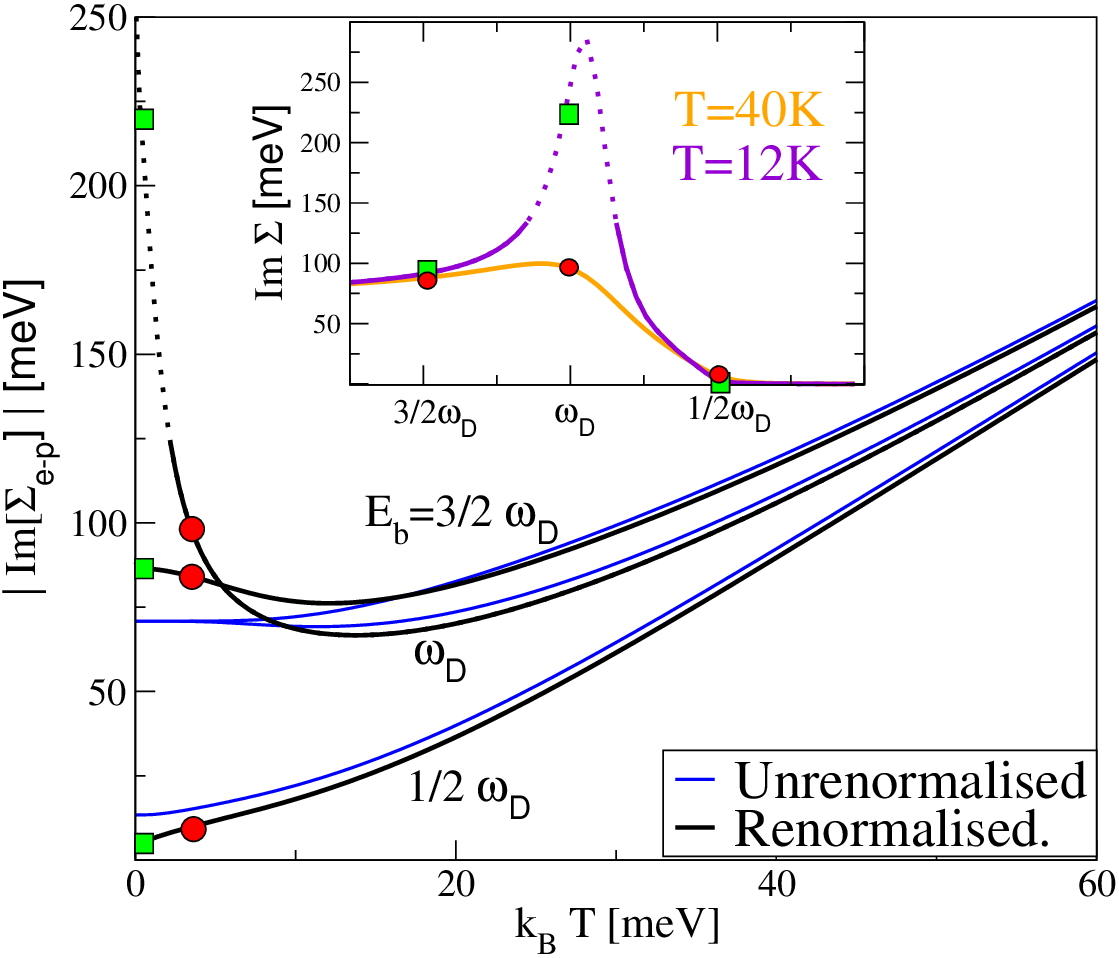}
\caption{\label{fig:Temp_dep} 
Temperature dependence of the broadening
at $\omega_{D}/2$, at $\omega_{D}$ and
at $3/2 \omega_{D}$ binding energies. Black lines represent the renormalised broadening
according to Eq. (\ref{renorm}).
Note the minimum at temperatures $k_{B} T \sim 13 \ meV$.
Black dotted line for $\omega_{D}$ indicates region where
Eq .\ref{renorm} is not reliable. The inset shows the 
$Im\Sigma$ dependence on binding energy ($E_b$).}
\end{figure}
Ref. \cite{Balasu1} reported
high temperature measurements of the broadening for 
the surface state hole at biding energy $\sim 350$ meV, obtaining 
an estimate $\lambda_{0}$(E=350 meV) of $1.15 \pm 0.1$ , which
compares reasonably well with a calculated $\lambda_{0}$(E= 350meV) $\simeq 0.9$ 
(see bottom panel of Fig. \ref{fig:ReS-ImS}). 

Analysing the temperature dependence of the broadening in more detail,
we note that for larger T , the maximum value of the T-derivative 
of $Re\Sigma$ decreases, and so does
the effect of the renormalising denominator in Eq. 6
(Fig. \ref{fig:Renor_ImS}).
By contrast, increasing temperature favours phonon occupation:
thus we have two competing tendencies and a minimum 
can be expected in the temperature dependence of $Im\Sigma$. 
Detailed calculation of $Im\Sigma$ predicts
a minimum of $Im\Sigma$ around $T \sim 150K$ for states close 
to $\omega_{D}$ binding energy (see Fig. \ref{fig:Temp_dep}). 
It would be interesting to verify this minimum in the temperature 
dependence of the broadening. This effect
should be even more pronounced for binding energies close 
to $\sim \omega_{D}$ where the effect of the renormalisation is stronger.
Here in fact the group velocity selectively rises 
{\em above} its bare value, and so does the
low temperature hole decay rate, giving rise to a local maximum as a
function of hole energy. Removal of this effect by
temperature causes the local minimum predicted in Fig 5.
%

In summary we presented the first  parameter-free ${\it ab-initio}$ calculation
of the e-p interaction in a surface state on a real metal surface,  Be(0001). 
The surface state renormalisation and its
broadening were calculated, and the latter was found to agree 
well with existing experimental data. The energy derivative of $Re\Sigma$ 
presents a strong energy and temperature dependence near $E_{F}$ , so that
an accurate measurement
of $\lambda \equiv \Lambda(E_{F}) =\lambda_{0}(E_{F})$
requires temperatures $T \lesssim 12K $ and 
energies very close to $E_F$. Measurements
done under different conditions can produce very different 
$\lambda$ values which explains the variety of
values in literature.
Additionally, the effect of the renormalisation 
of the surface state band on its own broadening is pointed out and 
new experiments are suggested to confirm a new minimum
predicted in the temperature dependence. Other conceivable consequences
of a medium size surface state e-p coupling, including 
the possible stabilisation of a 2D BCS state at T=0 \cite{Tosatti}
and/or the factors hindering it, remain to be explored.

We acknowledge financial support from the Basque Government,
the Max Planck Research Award Funds, the Spanish Ministerio
de Educaci\'on y Ciencia (MEC), and the Basque Country University.
Work in SISSA was supported through MIUR COFIN and FIRB projects, by
INFM/F and by the 'Iniziativa Trasversale Calcolo Parallelo' of INFM.


\begin{thebibliography}{30}

\bibitem{Bart85}
R. A. Bartynski {\it et al.},
Phys. Rev. B {\bf 32}, 1921 (1985).

\bibitem{Chul87}
E. V. Chulkov {\it et al.},
Surface Sci. {\bf 188}, 287 (1987).

\bibitem{general}
E. W. Plummer and J. B. Hannon, Prog. Surf. Sci. {\bf 46}, 149 (1994);
L. I. Johansson, {\it et al.}, Phys. Rev. Lett. {\bf 71}, 2453 (1993);
P. T. Sprunger {\it et al.}, Science {\bf 275}, 1764 (1997).

\bibitem{Balasu1}
T. Balasubramanian {\it et al.}, 
Phys. Rev. B {\bf 57}, R6866 (1998).

\bibitem{Balasu2}
S. LaShell {\it et al.},
Phys. Rev. B {\bf 61}, 2371 (2000).

\bibitem{Hengsberger-prl}
M. Hengsberger {\it et al.}, 
Phys. Rev. Lett. {\bf 83} 592 (1999).

\bibitem{Hengsberger-prb}
M. Hengsberger {\it et al.}, 
Phys. Rev. B {\bf 60}, 10796 (1999).

\bibitem{Eiguren}
A. Eiguren {\it et al.}, 
Phys. Rev. Lett. {\bf 88}, 066805 (2002).

\bibitem{EELS2}
J. B. Hannon {\it et al.},
Phys. Rev. B {\bf 53}, 2090 (1996).

\bibitem{Baroni}
S. Baroni {\it et al.},
Rev. Mod. Phys. {\bf 73}, 515 (2001). 

\bibitem{pwscf}
S. Baroni {\it et al.}, 
{\tt http://www.pwscf.org}.

\bibitem{Gironcoli2}
M. Lazzeri {\it et al.}, 
Surf. Sci. {\bf 402} 715 (1998). 


\bibitem{Grimvall81}
G. Grimvall,  in {\em The Electron-Phonon Interaction in Metals}, {\em Selected
Topics in Solid State Physics}, edited by E. Wohlfarth (North-Holland, New
York, 1981).

\bibitem{Tosatti} E. Tosatti, in Electronic surface
and interface states on metallic systems, eds, E. Bertel and M.
Donath (World Scientific, Singapore, 1995), p.67.


\end{thebibliography}
\end{document}